     \documentstyle[prb,twocolumn,aps,epsfig]{revtex}
     \newcommand{\be}{\begin{equation}}
     \newcommand{\ee}{\end{equation}}
     \newcommand{\bea}{\begin{eqnarray}}
     \newcommand{\eea}{\end{eqnarray}}

\begin{document}
     \draft
     \wideabs{
     \title{ Crossing of Specific Heat Curves in Liquid Helium 3 and Heavy
             Fermion Systems}
     \author{Suresh G. Mishra and P.A. Sreeram}
     \address{Institute of Physics, Sachivalaya Marg,
     Bhubaneswar 751 005, India.} 
     \maketitle
    \begin{abstract}

     Specific heat curves for various pressures, in many correlated electron 
     systems, have been seen to cross at a point. We analyze this behavior 
     using the spin fluctuation theory. It is found that when the system is
     considered near a ferromagnetic instability, the curves cross at a point
     and for systems near an antiferromagnetic instability, they cross at
     two points. The crossing behavior is related to crossover of these
     systems from quantum to classical fluctuation regimes. For detailed
     analysis, a weak linear pressure dependence of the crossover scale 
     $\alpha (0) T_F$ is assumed.

    \end{abstract} 
    \narrowtext
    }
     \vspace{.1in}

     Helium-3 is a Fermi system with a degeneracy temperature $\sim $
     5 K. However, in the normal phase it behaves 
     like a dense classical liquid for T $\geq $ .5 K and like a
     degenerate Fermi liquid below 0.2 K. Moreover, the (nuclear)
     spin susceptibility of $^3$He varies between 10 - 25 times the
     free Fermi gas susceptibility, $ \chi_P $, depending on
     pressure. The  coefficient $ \gamma $ of the linear
     term in specific heat is also large. There have been theoretical
     attempts to explain the low temperature behavior of $^3$He from
     two points of view phenomenologically. According to the first
     point of view, known as the spin fluctuation theory, the liquid
     is regarded as if it is near a ferromagnetic instability because
     of the largeness of $ \chi $. In this theory the temperature
     variation of various physical quantities is governed by
     transverse and longitudinal spin fluctuations. Though the actual
     transition does not take place, the effect of fluctuations is
     observable over a wide temperature range at low temperatures.
     This theory explains the temperature variation of many physical
     properties like spin susceptibility, specific heat
     quantitatively over a wide temperature range.\cite{MR78,MR85}
     In the second point of view one considers $^3$He as a liquid in
     the vicinity of a liquid to solid or the Mott-Hubbard
     ``metal-insulator'' transition. The reason being
     that the velocity of sound is large in liquid $^3$He. The
     compressibility at low temperatures is almost same as the
     compressibility of the solid phase. The liquid becomes sluggish
     at low temperatures before finally becoming a superfluid. This
     approach has been successful for understanding the pressure
     dependence of many properties.\cite{voll84}   

     In a recent publication \cite{Voll97} Vollhardt has drawn
     attention to intersection at a point of specific heat $ C_V(T) $
     curves for liquid helium 3 and other correlated Fermi systems.
     In the present work we want to emphasize that this behavior,
     particularly for $ ^3 $He, can be understood within the spin
     fluctuation theory. The formalism has been developed in detail
     earlier. In the following we use some results from
     \cite{MR78,MR85,SGM98} to discuss the crossing point in the specific 
     heat curves.  
     
     The spin fluctuation contribution to the free energy within the
     mean fluctuation field approximation (or quasi harmonic
     approximation) is given by,\cite{MR85}
     \begin{equation}
     \Delta\Omega =\frac{3 T}{2}\sum_{q,m}\ln \{ 1-U\chi_{qm}^0
     +\lambda T \sum_{q^\prime,m^\prime}D_{q^\prime m^\prime} \}.
     \label{freeenegy1}
     \end{equation}
     Where $D_{q,m}$ is the fluctuation propagator which is related
     to inverse dynamical susceptibility, $\chi_{qm}^0$ is the
     free Fermi gas (Lindhardt) response function, and $\lambda$ is the
     fluctuation coupling constant. The argument of
     the logarithm is related to inverse dynamic susceptibility.
     Considering only the thermal part of the integral and ignoring
     the zero point part, we perform the frequency summation and
     obtain, 
     \begin{equation}
     \Delta\Omega_{Thermal}=\frac{3}{\pi}\sum_q \int_0^\infty 
     \frac{d\omega}{e^{\omega/T}-1} \arctan \{\frac{\pi\omega/4q}
     {\alpha(T)+\delta q^2}\},
     \label{freeenegy2}
     \end{equation}
     Integrating over frequency, we get,
     \begin{equation}
     \Delta\Omega_{Thermal}=3 T \sum_q 
     \biggl( \ln\Gamma(y)-(y-\frac{1}{2})\ln(y)
     +y -\frac{1}{2}\ln (2\pi)\biggr).
     \label{freeenegy3}
     \end{equation}
     where, $ y = q (\alpha(\tau) + \delta q^2) / (\pi^2 \gamma \tau) $
     with $ \gamma = 1/2 $, $ \delta = 1/12 $, $ \tau = T/T_F $, the
     wavevector $ q $ is given in units of Fermi momentum $ k_F $ and
     the energy is in units of Fermi energy. Once the free energy
     correction is known, the specific heat correction is given by
     \begin{eqnarray} 
     \frac{\Delta C_v}{k_B} & = &
     -T\frac{\partial^2\Delta\Omega}{\partial T^2} 
     \nonumber \\
     & = & - 3T^2\sum_q \biggl[(\frac{2}{T}\frac{\partial y}{\partial
     T}
     +\frac{\partial^2 y}{\partial T^2})\phi(y)+(\frac{\partial
     y}{\partial T})^2
     \frac{\partial\phi(y)}{\partial y} \biggr] \nonumber \\
     & = & 6 \int q^2 dq \{\phi^\prime(y)(\frac{q}{\pi^2\gamma}
     \frac{\partial\alpha(T)}{\partial T}-y)^2 \nonumber \\
      & & + T \phi(y)
     \frac{q}{\pi^2\gamma} \frac{\partial^2\alpha(T)}{\partial T^2}\}.
     \label{ferro3dspe1}
     \eea
     The function $ \phi (y) $ is related to the fluctuation self
     energy summed over frequency. It varies as $ 1/2y $ for 
     $ y \ll 1 $ and as $ 1/12y^2 $ for $ y \gg 1 $. 
     
     Clearly the calculation of specific heat correction involves the
     temperature dependence of spin susceptibility. A self consistent
     equation for the temperature dependence of the inverse
     susceptibility (in units of $ \chi_P $) within one spin
     fluctuation approximation is given by,\cite{MR78,SGM98}
     \be
     \alpha(\tau) = \alpha(0) + \frac{\lambda}{\pi}\sum_q 
     q \phi(y) 
     \label{ferroalpha1}
     \ee
     For a finite $\alpha(0)$ there are two regions of temperature
     \cite{MR78}. For $ \tau < \alpha(0)$, which corresponds to $ y
     \gg 1 $, one gets 
     an enhanced Pauli susceptibility with standard paramagnon theory
     corrections, $ \alpha(\tau ) = \alpha (0) + x \tau^2 / \alpha
     (0) $, where $x$ turns out to be $\approx 0.44$. For $ \alpha(0)
     < \tau < 1 $, $ \alpha (\tau ) \sim \tau^{n} $ with the exponent
     $ 1 \le n \le 4/3 $. This result for the susceptibility in this
     regime mimics the classical Curie Weiss susceptibility. Hence
     even for a Fermi system for $ T < T_F $, the susceptibility
     behaves like the one for a collection of classical spins. This
     behavior agrees well \cite{MR78} with experimental results of
     Thompson et. al. \cite{Thompson70}. 

     In the low temperature limit ($ y \gg 1 $) it turns out that $
     \alpha ^{\prime} $ and $ \alpha ^{\prime \prime } $ do not
     contribute to the leading temperatutre dependence. The specific
     heat correction is given by,
     \be
     \frac {\Delta C_v }{k_B} = - \sum_q \frac {\pi^2 \tau } {4
     q(\alpha + \delta q^2 )}. 
     \ee
     The phase space integral reproduces the standard paramagnon mass
     enhancement result, $ \tau \ln \alpha $ for $ 
     \Delta C_v $,. The higher order terms give a $ (\tau ^3 /
     \alpha (0)^3) \ln (\alpha (0) /\tau^2 ) $.
          
     In the classical regime, $\alpha (0) \le \tau \ll 1 $ the small
     $y$ approximation holds, and $ \alpha (\tau ) $ varies as $ \tau
     $. In Eq. \ref{ferro3dspe1} $ \alpha ^{\prime \prime} = 0 $ and
     $ \alpha ^{\prime} = \alpha / \tau $, leading to $ \Delta C_v $
     falling as $ 1/\tau^2 $ at higher temperatures.  
     
     The result is that similar to the susceptibility variation 
     there are two regimes for the specific heat also and the
     behavior of the specific heat in these two regimes is
     qualitatively different. At
     low temperature there is an enhanced linear rise leading to a
     peak around 0.15 K and thereafter a slow fall as the temperature
     increases. This peak marks a transition from quantum to
     classical spin fluctuation regimes. Fig.\ref{expferro} shows a
     set of curves of $(C_v(P,T)-C_v(0,T))/C_v(0,T)$ for P = 15  Bar
     to P = 30 Bar. To calculate the specific heat, the free electron part 
     ($\pi^2 T/2 T_F$) has been added to $\Delta C_v(T)$. The value of 
     $\alpha(\tau)$ has been calculated self 
     consistently using Eqn.\ref{ferroalpha1} and then used as an input 
     in the specific heat calculation. The coupling constant $\lambda$ has
     been chosen to be $0.08$ and the cutoff for the momentum sum,
     $1.2k_F$. The crossover temperature is related to $
     \alpha (0) T_F $ which depends on pressure. 

 \begin{center}
      \epsfig{file=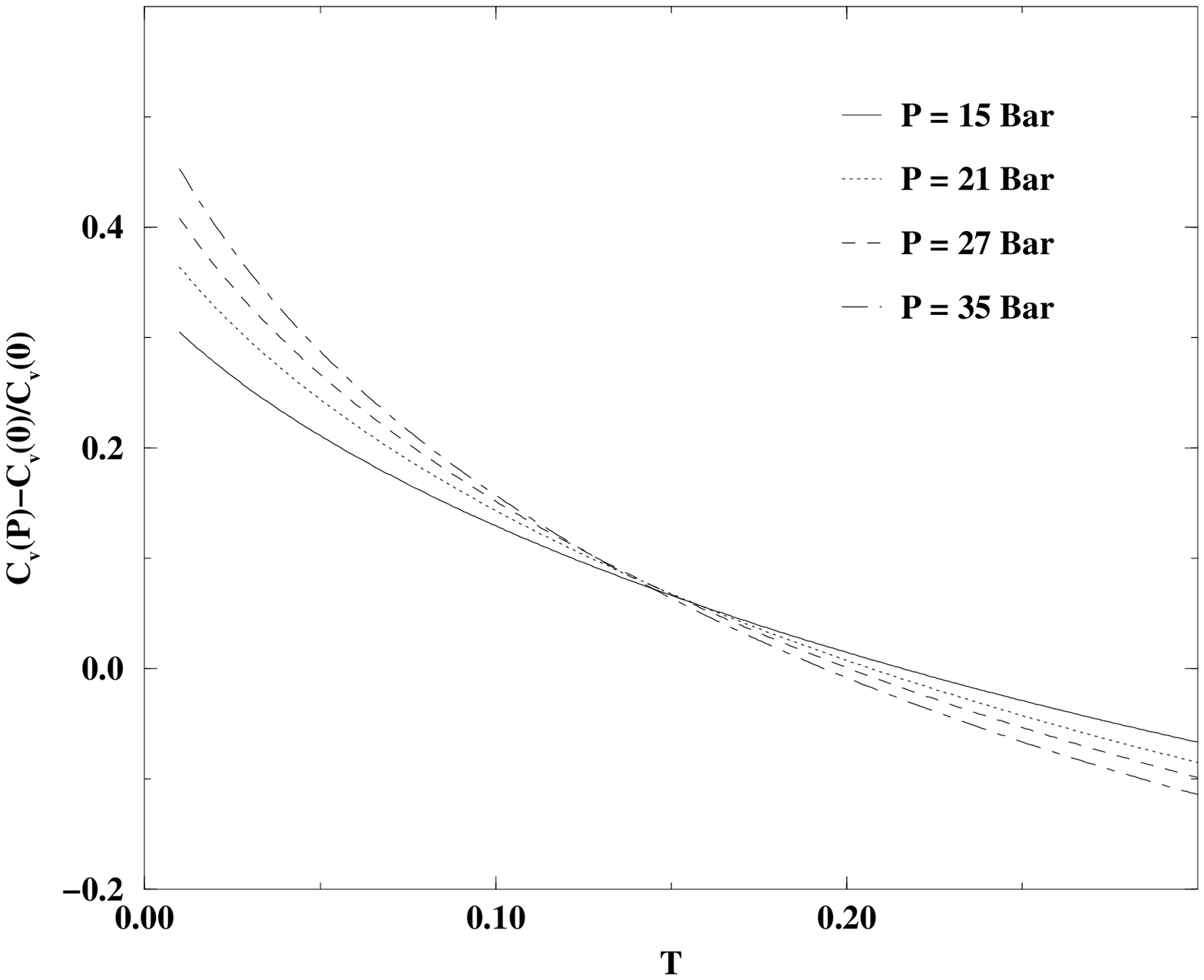,width=9cm,height=7cm}
      \vspace{-1.0cm}
      \begin{figure}[p]
      \caption{ $(C_v(P,T)-C_v(0,T))/C_v(0,T)$ as a function of $T$
      for P = 15, 21, 27 and 34.36 Bar respectively. The values of
      coupling constant $\lambda$ and momentum cutoff are chosen to
      be 0.08 and 1.2$k_F$ respectively. The curves for pressures
      below 15 Bar cross slightly away from the point shown in the curve
      for reasons mentioned in the text below.
      \label{expferro}}
      \vspace{0.5cm}
      \end{figure}
 \end{center}
 \begin{center}
      \epsfig{file=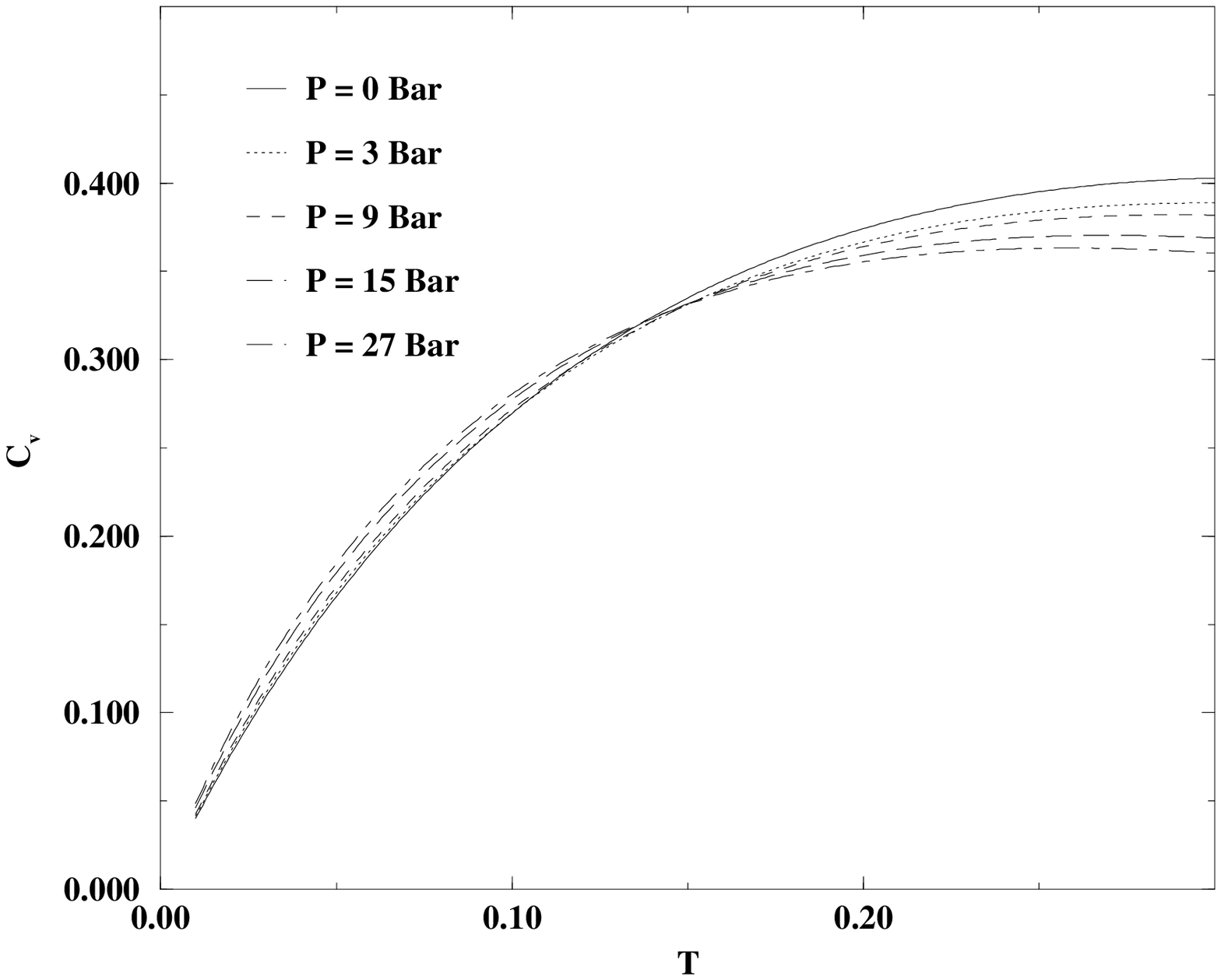,width=9cm,height=7cm}
      \vspace{-1.0cm}
      \begin{figure}[p]
      \caption{ $(C_v(P,T)$ as a function of $T$
      for P = 0, 3, 15, and 27 Bar respectively. The values of
      coupling constant $\lambda$ and momentum cutoff are chosen to
      be 0.08 and 1.2$k_F$ respectively. 
      \label{thferro}}
      \vspace{0.5cm}
      \end{figure}
 \end{center}

     It turns out that for $ \alpha (0) T_F $ scaling linearly with
     pressure the specific heat curves for various pressures cross at
     a point. The linear scaling is experimentally observed above 
     pressures about 15 kbar. However, at small pressures there is some
     departure.  In Fig.\ref{thferro}, $C_v(P,T)$ is plotted as a function 
     of $T$ for various values of P, assuming a linear dependence of
     $\alpha (0) T_F $ on pressure. It is very clear from the figure that 
     for linear scaling of $\alpha (0) T_F $ the curves cross at a point.
     The crossing point increases very slightly with
     increase in cutoff and with decrease in $\lambda$ but the nature
     of crossing is not affected. The crossing of the specific heat
     curves at various pressures has also been discussed in the
     earlier work of Seilers et. al. \cite{Seil86} but they do not match 
     with
     Greywall's experimental findings of crossing at a point\cite{Grey83}, 
     in fact
     there is wide range of temperatures over wich the curves cross.
     In the present spin fluctuation calculation with the assumption
     of a linear scalling the crossing occurs at a point. 
     
     We have used the terms quantum and classical in the discussion
     above, because, temperatures below $\alpha (0) T_F $ essentially
     define a regime where one gets a Fermi liquid behavior whereas
     at the higher temperatures, fluctuations get correlated
     resulting in the classical behavior for susceptibility and the
     specific heat corrections.  The distinction, quantum versus
     classical, becomes clear when one takes the limit $ \alpha (0)
     \rightarrow 0 $. In that case the Curie law is obtained down to
     zero degree,\cite{SGM98} while in the opposite limit ($\alpha
     (0) \rightarrow 1 $) one gets the Pauli susceptibility. The
     crossing temperature $ T_{+} $ is related to the pressure
     derivative of this this crossover scale. This crossover
     characterizes the change in temperature dependence of all
     thermodynamic properties.  

 \begin{center}
      \epsfig{file=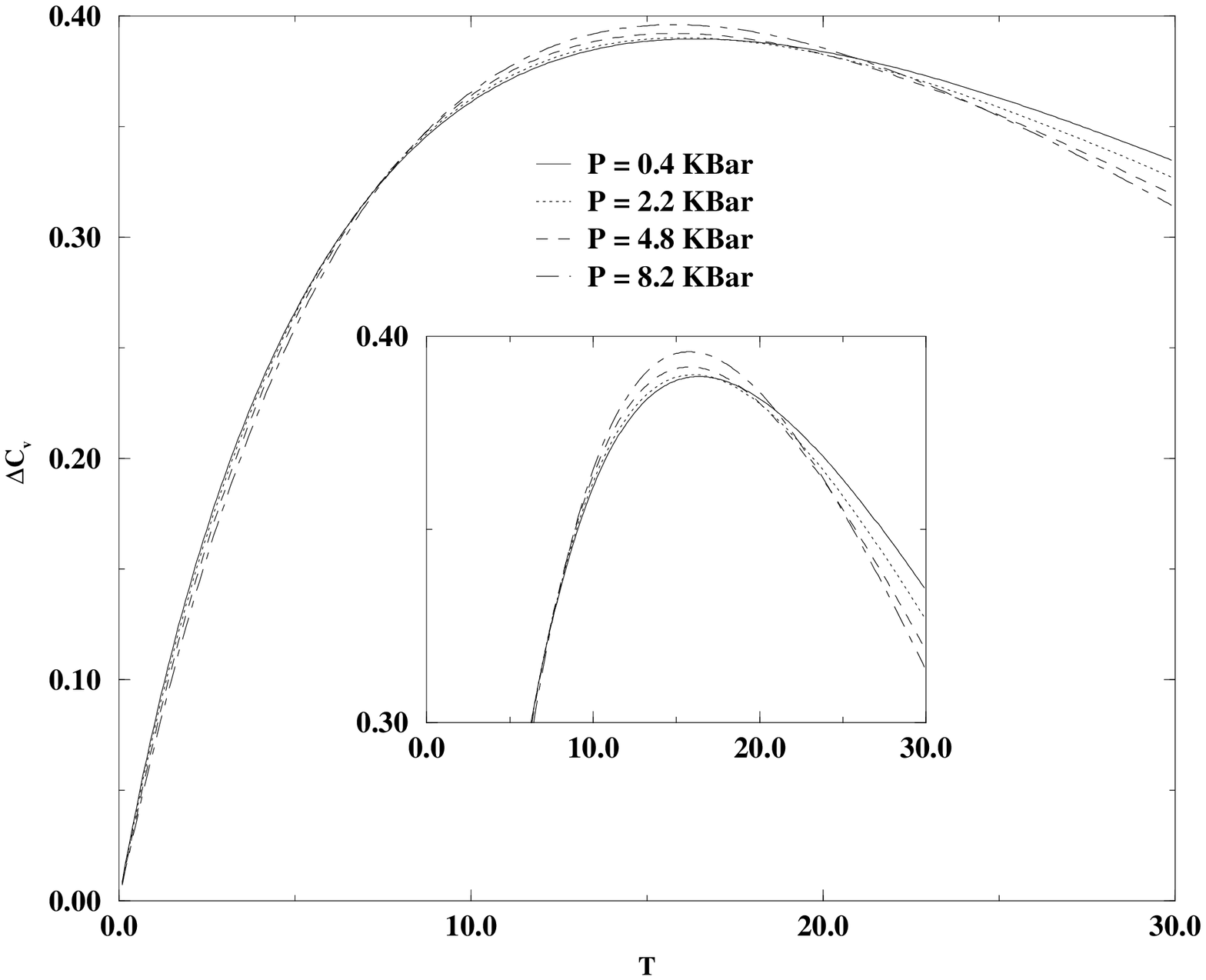,width=9cm,height=7cm}
      \vspace{-1.0cm}
      \begin{figure}[p]
      \caption{ $\Delta C_v(P,T)$ as a function of $T$
      for P = 0.4, 2.2, 4.8 and 8.2 KBar respectively. The values of
      coupling constant $\lambda$ and momentum cutoff are chosen to
      be 1.5  and 1.2$k_F$ respectively. Inset figure shows the region
      close to the crossover points. 
      \label{thanti}}
      \vspace{0.5cm}
      \end{figure}
 \end{center}

     There are some heavy fermion materials, for example
     CeCu$_{6-x}$Au$_x$ \cite{Ger89}, CeAl$_3$ \cite{Bro86} in which
     the specific heat curves cross. However, the crossing occurs at
     two points, in case of 
     CeAl$_3$ these temperatures are 5K and 17K respectively. It is
     possible to cast the behavior of these materials in terms of
     spin fluctuation theory for antiferromagnets. We have calculated
     the specific heat corrections by writing the equations for the
     susceptibility enhancement and specific heat near an
     antiferromagnetic instability.\cite{SGM98} The curves do cross
     at two points as shown in the Fig\ref{thanti}. The detailed comparison 
     with experiments is in progress and will be presented elsewhere.

\end{document}